\documentstyle[epsfig,float]{aipproc}

\begin{document}
\title{Chiral Symmetry Amplitudes in the S-Wave Isoscalar and
Isovector Channels and the $\sigma, f_0(980), a_0 (980)$
SCALAR MESONS}
\author{J.A. Oller and E. Oset}

\address{Departamento de F\'{\i}sica Te\'orica and IFIC\\
Centro Mixto Universidad de Valencia-CSIC\\
46100 Burjassot (Valencia), Spain}

\maketitle

\begin{abstract}
We use a nonpertubative approach which combines coupled channel Lippmann
Schwinger equations with meson-meson potentials provided by the lowest
order chiral Lagrangian. By means of one parameter, a cut off in the
momentum of the loop integrals, which results of the order of 1 GeV, we
obtain singularities in the S-wave amplitudes corresponding to the 
$\sigma$, $f_0$
and  $a_0$
resonances.
The  $\pi \pi
\rightarrow \pi \pi \, , \, \pi \pi \rightarrow
K \bar{K}$  phase shifts and inelasticities in the $T = 0$ scalar
channel are well reproduced as well as the $\pi^0 \eta$
and $K \bar{K}$ mass distributions
in the $T = 1$ channel. Furthermore, the total and partial 
decay widths
of the $f_0$  and $a_0$ resonances are properly reproduced. 
The results seem to indicate that chiral symmetry
constraints at low energy and unitarity in coupled channels is the basic 
information contained in the meson-meson interaction below $\sqrt{s}
\simeq  1.2 \; GeV$.
\end{abstract}

\section{Introduction}

The understanding of the meson-meson interaction in the scalar sector
is still problematic. There is some debate about the spectrum of hadronic 
states and even more about its nature. Below $\sqrt{s} = 1.2 \;
GeV$, which we will study here,
the existence of a broad scalar-isoscalar meson around $500 \; MeV$ has had
permanent ups and downs. However, the $f_0 (980)$  (also called $S^*$, 
$I^G (J^{PC}) = 0^+ (0^{++}))$,  and the $a_0 (980)$ (also called $\delta$, 
$I^G (J^{PC}) = 1^- (0^{++})) $
mesonic states are well established experimentally, although there is still
debate around their decay widths and particularly about their nature.

\section{$L= 0,  T= 0,1$ strong amplitudes in the coupled channel 
approach}

We assume that the lowest order Hamiltonian provides us with the potential 
that we iterate in the Lippmann Schwinger equation with two coupled channels, 
using relativistic meson propagators in the intermediate states. Our channels 
are labelled 1 for the $K \bar{K}$ and 2 for the $\pi \pi$ states in $T=0$ 
and 1 for $K \bar{K}$, 2 for $\pi \eta$ in  $T = 1$. 
\begin{equation}
{\cal L}_2 = \frac{1}{12 f^2} < ( \partial_\mu
\Phi \Phi - \Phi \partial_\mu \Phi)^2 + M \Phi^4 >
\end{equation}
\vspace{-0.5cm}
\begin{equation}
\begin{array}{ll}
t_{ij}  =  V_{ij} + \sum_k V_{ik} G_{kk} t_{kj} &
G_{ii} = i \frac{1}{q^2 - m_{1i}^2 + i \epsilon} \; \; 
\frac{1}{(P - q)^2 - m_{2i}^2 + i \epsilon}
\end{array}
\end{equation}
where $P$ is the total fourmomentum of the meson-meson systems and $q$ the 
fourmomentum of one intermediate meson. The former equation is, of 
course, an integral one.The loop integral in eq. (2) is divergent. In our 
approach we take a cut off $q_{max}$ ($ \Lambda=\sqrt{m_k^2+q_{max}^2} $) 
for the maximum value of the modulus of 
the momentum $q$. On grounds of chiral symmetry computations we should expect 
this cut off to be around $1 \, GeV$ \cite{36}.By fixing the cut off one 
generates the counterterms in our scheme.

Let us see that only on shell information is needed. The argument goes as 
follows: We can write the off shell amplitudes coming from ${\cal L}_2$ as 
\begin{equation}
\begin{array}{c}
V=V_{on}+\beta \sum_i (p_i^2-m_i^2) \\
V^2=V^2_{on}+2 \beta V_{on} \sum_i (p_i^2-m_i^2)+\beta ^2\sum_{ij}
(p_i^2-m_i^2)(p_j^2-m_j^2)
\end{array}
\end{equation}
When performing the $q^0$ integration in the loop we have two poles.Let us take the 
contribution from the first pole (the procedure follows analogously for the 
second pole). From the second term in eq. (3) we get the contribution
\begin{equation}
\frac { 2 \beta V_{on} } { (2 \pi )^3 } \int \frac { d^3q }{ 2w_1(q) } \frac
 { (P^0-w_1)^2-w_2^2 } { (P^0-w_1)^2-w_2^2 }
=\frac { \beta V_{on} } { 2 \pi^2 } \int dw_1 q
\end{equation}
which for large $\Lambda$ compared to the masses goes as $V_{on} \Lambda ^2$
 and has the same structure in the dynamical variables as the tree 
diagram. The same happens for the rest of off shell terms, that is, they  
 combine with 
the tree level contribution, giving rise to an amplitude with the same 
structure as the tree level one but with renormalised parameters $f$ and 
masses. However, since we are taking physical values for $f(f=f_\pi=93 \, MeV
)$ and the masses 
in the potential, these terms should be omitted. One can proceed like that 
to higher orders with the same conclusions. Since we are taking V and $t$ 
on shell they factorize outside the $q$ integral. Thus the term VGT, after
the $q^0$ integration is performed, is given by
\vspace{-0.1cm}
\begin{equation}
\begin{array}{c}
V_{ij} G_{jj} t_{jk} = V_{ij}
(s) t_{jk} (s) G_{jj} (s) \\
G_{jj} (s) = \int_0^{q_{max}} \frac{q^2 dq}{(2 \pi)^2}\frac{\omega_1 + 
\omega_2}{ \omega_1 \omega_2 [P^{02} - (\omega_1 + \omega_2)^2 + i 
\epsilon]}
\end{array}
\end{equation}
\vspace{-0.1cm}
where $\omega_i = (\vec{q}\,^2 + m_i^2)^{1/2}$ 
and $P^{02} = s$
and the subindex $i$ stands for the two intermediate mesons of the $j$
channel.Thus the coupled channel Lippmann Schwinger equations get reduced to
a set of algebraic equations which are solved in a trivial way. Note also 
that $V_{12} = V_{21}$ and $t_{12}=t_{21}$ by time reversal invariance.
\begin{equation}
\begin{array}{ll}
\Delta_\pi = 1 - V_{22} G_{22} & 
\Delta_K = 1 - V_{11} G_{11} \\ 
\Delta_c = \Delta_K \Delta_\pi - V_{12}^2 G_{11} G_{22} &
\Delta = det A = \Delta_\pi \Delta_c
\end{array}
\end{equation}
\vspace{-0.5cm}
\begin{equation}
\begin{array}{ll}
t_{11} = \frac{1}{\Delta_c} (\Delta_\pi V_{11} + V_{12}^2 G_{22}) &
t_{21} = \frac{1}{\Delta_c} (V_{21} G_{11} V_{11} + \Delta_k V_{21})\\
t_{22} = \frac{1}{\Delta_c} (\Delta_K V_{22} + V_{12}^2 G_{11}) 
\end{array}
\end{equation} 

The physical amplitudes as a function of $s$ (real variable) are given by
the expressions in eq. (7). However, in order to explore the position 
of the poles of the scattering amplitudes one must take into account
the analytical structure of these amplitudes in the different Riemann
sheets. These sheets appear because of the cuts related to the opening
of thresholds in $G_{jj} (s)$  (see eq. (5)). 

\section{Results}

\begin{figure}[h]

\centerline{
\protect
\hbox{
\psfig{file=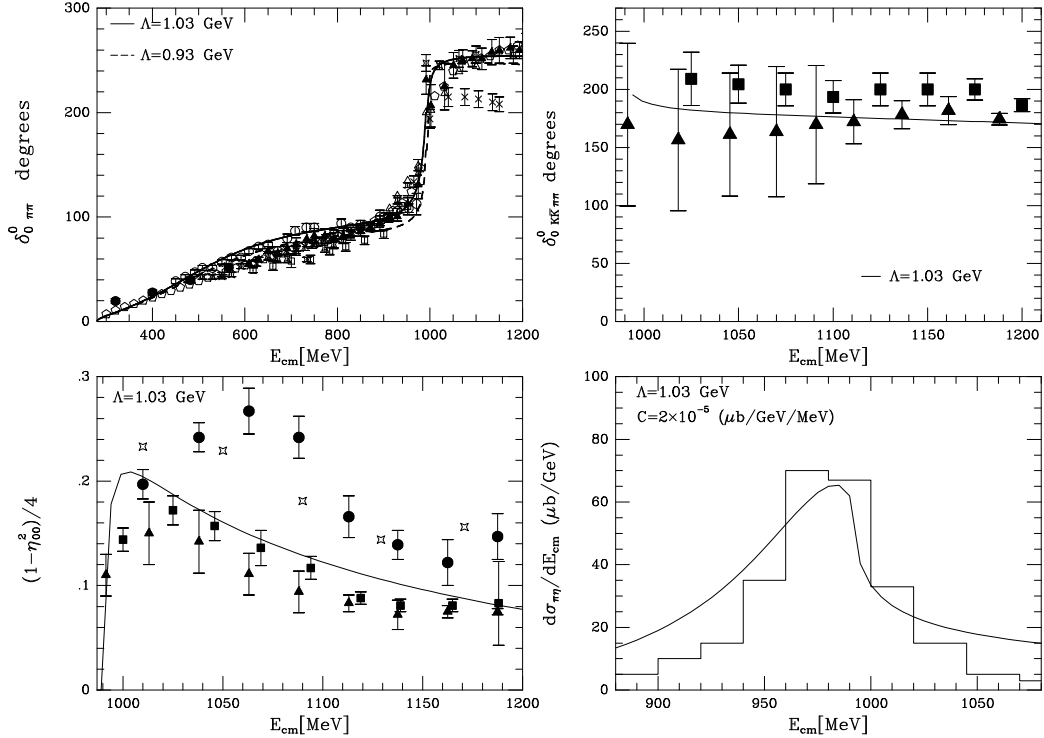,width=0.7\textwidth,angle=-90}}}
\caption{I=0, phase shifts for $\pi\pi\rightarrow \pi\pi$, $K\bar{K}
\rightarrow \pi\pi$ and inelasticities. I=1, mass distribution of $\pi^-\eta$ 
in $K^- p \rightarrow \sum^+ (1385) \pi^- \eta$}

\end{figure}

$
\begin{array}{ll}
\begin{tiny}
\begin{minipage}[t]{0.5\textwidth}
\begin{center}
{\small{Table I: $f_0$ Mass and partial widths ($\Lambda = 1.03 \, GeV$)}}
\end{center}

\begin{center}
\begin{tabular}{|c|c|l|}
\hline
$f_0 $ & our results & $\quad$ experiment\\
\hline
$\begin{array}{c}
\hbox{Mass}\\
\hbox{Pole position}\\
 \, [MeV] \end{array}$  & 993.5 & 980 $\pm$ 10 \\
\cline{1-2}
$\begin{array}{c}
\hbox{Mass} \\
\hbox{Peak of the} \, t_{21}\\ 
\hbox{amplitude}\\
\, [MeV] \end{array}$ & 982 &  980 $\pm$ 10 \\
\hline
$\Gamma_{tot}$ [MeV] & 25 & 40 - 100 \\
\hline
$ \frac {\Gamma_{\pi \pi}}{\Gamma_{\pi \pi} + \Gamma_{K \bar{K} } }$ & 0.68 & 
$\begin{array}{l}
0.67  \pm  0.009  \;  \\
0.81 \pm  _{0.009} ^{0.004} \;  \\
0.78  \pm  0.003 \;  \\
\hbox{(av.)} \, 
0.781  \pm  0.024 \;  
\end{array}$\\
\hline
$\begin{array}{c}
\Gamma_{\gamma \gamma}\\
(KeV) \end{array} $ & 0.2 &
$\begin{array}{l}
0.63 \pm 0.14 \, \\
0.42 \pm 0.06 \pm 0.18 \, \\  
0.29 \pm 0.07 \pm 0.12 \, \\  
0.31 \pm 0.14 \pm 0.09 \, \\  
\hbox{(av)} \, 0.56 \pm 0.11  \,  \end{array}$\\  
\hline
\end{tabular}
\end{center}

\end{minipage}
\end{tiny}  &
\begin{tiny}
\begin{minipage}[t]{0.5\textwidth}
\begin{center}
{\small{Table 2: $a_0$ Mass and partial width ($\Lambda = 1.03 \, GeV$)}}
\end{center}
\begin{center}
\begin{tabular}{|c|c|l|}
\hline
$a_0 $ & our results & $\quad \; $experiment\\
\hline
$\begin{array}{c}
\hbox{Mass}\\
\hbox{Pole position}\\
 \, [MeV] \end{array}$  & 1009.2 & 983.5 $\pm$ 0.9 \\
\cline{1-2}
$\begin{array}{c}
\hbox{Mass}\\
\hbox{Peak of the} \, t_{21}\\ 
\hbox{amplitude}\\
\, [MeV] \end{array}$ & 982 &  983 $\pm$ 0.9 \\
\hline
$\Gamma_{tot}$ [MeV] & 112  & 50 - 100 \\
\hline
$\frac{\Gamma_{K \bar{K}}}{\Gamma_{\eta \pi}}$ & 0.60 & 
$\begin{array}{l}
1.16  \pm  0.18  \;  \\
0.7 \pm  0.3 \;  \\
0.25 \pm  0.008 \;  \\
\end{array}$\\
\hline
$\begin{array}{cl}
\frac{\Gamma_{\gamma \gamma} \cdot \Gamma_{\eta \pi}}{\Gamma_{tot}}\\
\, [KeV] \end{array} $ & 0.49 & $\begin{array}{l}
0.28  \pm  0.04 \pm 0.01  \;  \\
0.19 \pm  0.07 \pm _{0.1} ^{0.07} \; \\
\hbox{av.} \, 
0.24   \pm _{0.008} ^{0.007}  \;  \\
\end{array}$\\
\hline
\end{tabular}
\end{center}

\end{minipage}
\end{tiny}
\end{array}
$                                                                     
All these data can be obtained from ref. \cite{PDG}, PDG.

Our calculation for the $\gamma \gamma$ width of both resonances is 
explained in our other contribution to these proceedings, {\bf 
THEORETICAL STUDY OF THE $\gamma \gamma \rightarrow$ MESON-MESON 
REACTION}.

{\bf $\sigma$ pole}: 
$
\begin{array}{cc}
Mass: $469$ MeV & Width: $406$ MeV 
\end{array} 
$

\section{Conclusions}

We have used a non perturbative approach to deal with the meson meson
interaction in the scalar sector at energies below 
$\sqrt{s} \simeq 1.2 \; GeV$, exploiting 
chiral symmetry and unitarity in the coupled channels. The Lippmann
Schwinger equation with coupled channels, using relativistic meson 
propagators, and the lowest order chiral Lagrangians, providing 
the meson-meson potential, are the two ingredients of the theory. 
One cut off in momentum is also in order to cut the loop integrals
and it is fitted to the data \cite{Oller}. 
This is the only parameter of the theory. The best fit is found with 
$q_{max} \simeq 900 \; GeV$, or equivalently $\Lambda =1.03 \; GeV$. 
With only these elements we find a good agreement with the experiment 
for phase shifts, inelasticities and mass distributions until $\sqrt{s} 
\simeq 1.2$ $GeV$ generating dynamically the $f_0(980), a_0(980)$ 
and $\sigma$ resonances.

\end{document}